\documentclass{PoS}

\usepackage{amsmath}

\def\pp#1{\partial_{#1}}

\def\s[#1,#2]{[#1\stackrel{\star}{,}#2]}
\def\eq{\begin{equation}}
\def\en{\end{equation}}
\def\bit{\begin{itemize}}
\def\eit{\end{itemize}}

\title{Nambu Sigma Model and Branes}

\ShortTitle{Nambu Sigma Model}

\author{\speaker{Peter Schupp}\thanks{Current address: Maxwell Institute for Mathematical Sciences, Heriot-Watt University, Edinburgh, UK.}\\

        Jacobs University Bremen, 28759 Bremen, Germany\\
        E-mail: \email{p.schupp@jacobs-university.de}}

\author{Branislav Jur\v co\\
       Mathematical Institute, Charles University, Prague 186 75, Czech Republic\\
        E-mail: \email{jurco@karlin.mff.cuni.cz}}

\abstract{In analogy to Nambu's generalization of mechanics, we present a generalization of Poisson sigma models to higher dimensional world volumes.
We find corresponding generalizations of string sigma models and open-closed string relations for background fields to the case of $p$-branes.
As an application, we discuss a DBI-type effective action for open membranes.}

\FullConference{Proceedings of the Corfu Summer Institute 2011 School and Workshops on Elementary Particle Physics and Gravity\\
		September 4-18, 2011\\
		Corfu, Greece}

\begin{document}

\section{Introduction}
The effect of a closed string background on the physics of open strings has an elegant description involving spacetime non-commutativity. A system of open strings ending on a single D-brane is effectively described by the Dirac-Born-Infeld action and the effect of a closed string background is captured by an equivalent non-commutative version of that action.\footnote{The equivalence of commutative and non-commutative descriptions has originally been shown to be exact in the limit of slowly varying fields~\cite{Seiberg:1999vs}. This limiting assumption can be dropped and the equivalence becomes an identity at the level of the actions, when a semi-classical version of the underlying  non-commutative geometry is used~\cite{Jurco:2001my}.} The fields and parameters in the equivalent descriptions are related by certain open-closed string relations. Intuitively, one would expect similar phenomena in the case of open membranes in the presence of background fields. This is in fact the case, with non-commutativity replaced by higher geometric structures based on Nambu-Poisson brackets. Taking the generalization of classical mechanics by Nambu~\cite{Nambu:1973qe} as a guideline, we are led to $p$-brane analogs of Poisson sigma models, string sigma models, open-closed string relations, and the Dirac-Born-Infeld action. We propose the latter ``Nambu-Dirac-Born-Infeld action'' as a natural candidate for the bosonic part of an effective description for branes ending on branes. The strategy underlying this work was originally inspired by related ideas in an earlier work on Nambu-Poisson M5-brane theory~\cite{Chen:2010br}. For a more complete list of references and computational details, we refer to \cite{Jurco:2012yv}.

\section{Nambu mechanics}
 In this section, we will recall some basic facts about Nambu mechanics and Nambu-Poisson tensors and will then rewrite the pertinent axioms in a way that is suitable for our purposes.

Nambu mechanics~\cite{Nambu:1973qe,Takhtajan:1993vr} describes multi-Hamiltonian dynamics with generalized Poisson brackets. An elegant application are Euler's equations for the spinning top, which are usually written in a principal axis system for the angular velocities $\omega_1$, $\omega_2$, $\omega_3$:
\begin{equation}
  I_1 \dot \omega_1 + \omega_2 \omega_3 (I_3 - I_2) = 0\,, \quad
  I_2 \dot \omega_2 + \omega_3 \omega_1 (I_1 - I_3) = 0\,, \quad
  I_3 \dot \omega_3 + \omega_1 \omega_2 (I_2 - I_1) = 0\,.
\end{equation}
In terms of angular momenta $\vec L = I \cdot \vec \omega$ this becomes
\begin{equation}
\dot L_i = \epsilon_{ijk} L_j L_k / I_j =
\left\{L_i , T, \frac{1}{2}\vec L^2\right\} \,,
\end{equation}
where
 $T = \frac{1}{2} \vec L \cdot \vec \omega $ is the kinetic energy and we have introduced the Nambu-Poisson bracket
\begin{equation}
\{f,g,h\} = \det{\left[\frac{\partial(f,g,h)}{\partial(L_1,L_2,L_3)}\right]}
= \epsilon^{ijk} \,\partial_i f \,\partial_j g \,\partial_k h \,.
\end{equation}
More generally, Nambu-Poisson structures are generalizations of Poisson structures, with the Poisson bracket replaced by a Nambu-Poisson bracket, which is  a skew-symmetric multi-linear derivation
\begin{equation}
\{f,h_1,\ldots,h_p\} =  \Pi^{i\, j_1 \ldots j_p}(x) \,\partial_i f\, \partial_{j_1} h_1\, \cdots\, \partial_{j_p} h_p
\end{equation}
that satisfies the fundamental identity
\begin{eqnarray} \label{fundamental}
\{\{ f_0,\cdots,f_p\}, h_1,\cdots , h_p\} & = & \{\{ f_0, h_1,\cdots , h_p\},f_1,\cdots,f_p\}+ \ldots\nonumber \\
&&\ldots +\{f_0,\ldots,f_{p-1},\{ f_p, h_1,\cdots , h_p\}\} \,.
\end{eqnarray}
For $p=1$ this reduces to the Jacobi identity for the Poisson bracket, which implies a differential constraint on the Poisson tensor: $\theta^{il} \partial_l \theta^{jk} + \text{($i,j,k$ cyclic)} = 0$. For  $p>1$ the fundamental identity is a differential as well as a rather restrictive algebraic constraint that implies that $\Pi$ factorizes:
\begin{equation}
\Pi = V_0\wedge V_1 \wedge \ldots \wedge V_p \,,
\end{equation}
with $p+1$ suitable linearly independent vector fields $V_0$, \ldots $V_p$.
A Nambu-Poisson structure thus leads to a foliation of the underlying manifold into $(p+1)$-dimensional submanifolds that are determined by the
vector fields $V_\alpha$.  By local orthogonal transformations, the Nambu-Poisson tensor can be brought into the canonical form
\begin{equation}
\Pi^{i\, j_1 \ldots j_p}(x) = \left\{ \begin{array}{ll}
|\Pi(x)|^\frac{1}{p+1} \:\epsilon^{i\, j_1 \ldots j_p} & \text{ for } \quad i,j_1, \ldots, j_p \in \{0,1,\ldots,p\} \\
0 & \text{ else } \end{array} \right. \,,
\end{equation}
where $|\Pi(x)|$ is the generalized determinant of the (rectangular) matrix $\Pi^{i\, J}(x)$.

The three defining properties of a Poisson structure can be motivated from the point of view of physics as follows: The derivation property ensures well defined equations of motion with unique solutions given suitable initial conditions, the Jacobi identity ensures that time evolution is canonical in the sense that it preserves Poisson brackets, and skew symmetry of the Poisson bracket ensures that the Hamiltonian is a constant of motion (because it implies $\{H,H\} =0$). This point of view can be generalized to $p>1$, yielding a characterization of Nambu-Poisson structures, that deemphasizes the introducing of multiple Hamiltonians and is more suitable for our purposes: \begin{itemize}
\item
A Nambu tensor
$\Pi \in TM \otimes \Lambda^p TM$
maps a time-evolution $p$-form $\tilde\eta$ ``Nambuian'' to a time-evolution vector field
\begin{equation} \label{nambuian}
\Pi(\tilde\eta)  = \frac{1}{p!}\Pi^{i\, j_1\ldots j_p} \,\tilde\eta_{j_1\ldots j_p} \partial_i \equiv \Pi^{i J} \tilde\eta_J \partial_i \in TM \,.
\end{equation}
(In the last expression we have used ordered multi-index notation: $J = (j_1,\ldots,j_p)$ with $j_1 < \ldots < j_p$. For $p=1$ the Nambuian is related to the Hamiltonian $H$ by  $\tilde\eta = dH$ and the time evolution vector field is then the Hamiltonian vector field.)
\item Canonical transformation property:
\begin{equation}
d\tilde\eta = 0 \quad \Rightarrow \quad \mathcal L_{\Pi(\tilde\eta)} \, \Pi = 0
\end{equation}
\item Conservation law property:
\begin{equation}
\tilde\eta = dh_1 \wedge \ldots \wedge dh_p \quad \Rightarrow \quad  \mathcal L_{\Pi(\tilde\eta)} \, \tilde\eta =  0
\end{equation}
\end{itemize}
These three axioms are equivalent to the usual definition of a Nambu-Poisson structure that we have given earlier. It is tempting to drop the conservation law property, as it is least needed in the following.
In our first application of these structures, $\tilde\eta_J$ will be a $\Lambda^p T^*M$-valued world volume auxiliary field. In the application to generalized gauge theory and $p$-branes, $\Lambda$ is a $p$-form gauge transformation parameter and the vector field $\Pi(\Lambda)$ generates semi-classical generalized gauge transformations. This is in analogy to semi-classical non-commutative abelian gauge transformations, which can be interpreted as special types of canonical transformations.

\section{Poisson sigma model and string sigma model}

Poisson sigma models were introduced in the context of 2-dimensional gravity \cite{Ikeda:1993fh,Schaller:1994es}. They later played a major role in the solution of the problem of deformation quantization \cite{Cattaneo:1999fm}. The action of a Poisson sigma model describes an open string propagating on a Poisson manifold $(M,\Pi)$ (target space):
\begin{equation}
S_\Pi = \int_\Sigma \left( A_i \wedge dX^i - \frac{1}{2} \Pi^{ij} A_i \wedge A_j\right)\,, \qquad \Pi = \frac{1}{2} \Pi^{ij}(X)\pp i \wedge \pp j \,,
\end{equation}
with embedding functions $X: \Sigma \rightarrow M$ and a 2-dimensional world sheet $\Sigma$.
The fields $A(\sigma)$  are 1-forms on $\Sigma$ with values in $T^*_{X(\sigma)} M$.
The equations of motion are
\begin{equation}
d X^i - \Pi^{ij} A_j = 0 \,, \qquad
d A_i + \frac{1}{2} \pp i \Pi^{kl} A_k \wedge A_l = 0 \,.
\end{equation}
Consistency of the equations of motions requires
\begin{equation}
\left[\Pi, \Pi\right]_S^{ijk} = \frac{1}{3}\left(\Pi^{il} \pp l \Pi^{jk} + \text{cycl}\right) = 0\,,
\end{equation}
i.e.\ the bi-vector $\Pi$ must satisfy the Jacobi identity and $(M,\Pi)$ is indeed a Poisson manifold.

The term `Poisson sigma model' generally refers to the 2-dimensional topological field theory described above. Adding a metric term in the action we arrive at the non-topological generalized Poisson $\sigma$-model
\begin{equation} \label{genPoissonsigma}
S = \int_\Sigma \left( A_i \wedge dX^i - \frac{1}{2} \Pi^{ij} A_i \wedge A_j - \frac{1}{2} (G^{-1})^{ij} A_i \wedge *A_j \right) \,.
\end{equation}
(We are working with world sheet signature $(-,+)$ and volume form $d^2 \sigma = d\sigma^0 \wedge d\sigma^1$.)
The $A_i = A_{i\alpha}(\sigma) d\sigma^\alpha$, $\alpha = 0,1$, are auxiliary fields.  On shell, i.e.\ using the equations of motion for $A_i$, \eqref{genPoissonsigma} is equivalent to the string sigma model
\begin{equation}
S' = -\int_\Sigma \frac{1}{2} \left(g_{ij} dX^i \wedge *dX^j + B_{ij} dX^i \wedge dX^j\right) \,,
\end{equation}
where $g$ and $B$ are related to $G^{-1}$ and $\Pi$ by the closed-open string relations~\cite{Seiberg:1999vs}
\begin{equation} \label{close-open-string}
\frac{1}{g+B} = G^{-1} + \Pi \quad \Rightarrow \quad G = g - B g^{-1} B \,, \quad \theta = - G^{-1} B g^{-1} \,.
\end{equation}


We would like to generalize these models to the case of $1+p$ dimensional world volumes and target spaces with Nambu-Poisson structures.
In the construction, we shall be guided by the fact that Nambu-Poisson structures are naturally interpreted as maps from (target space) $p$-forms to vector fields.
Before proceeding, it is useful to write \eqref{genPoissonsigma} explicitly in terms of the components $\eta_i = \eta_i(\sigma)d\sigma^1 :=  -A_{i 1}(\sigma)d\sigma^1$ and $\tilde \eta_j = \tilde\eta_j(\sigma) d\sigma^0 := A_{j 0}(\sigma)d\sigma^0$:
\begin{equation} \label{etaPoissonsigma}
  S = \int_\Sigma \left( dX^i \wedge \eta_i + \tilde \eta_j \wedge dX^j  - \Pi^{ij} \tilde\eta_j \wedge \eta_i
  - \frac{1}{2} G^{ij} \eta_i\wedge *\eta_j -\frac{1}{2} G^{ij} \tilde \eta_i \wedge *\tilde \eta_j \right) \,.
\end{equation}

\section{Nambu sigma model and brane sigma model}

In this section we will describe a Nambu sigma model that generalizes Poisson sigma models  in a similar way as Nambu mechanics generalizes ordinary mechanics.

In view of our preceding discussion, a natural generalization of the Poisson sigma model to the case $p>1$ will involve $p$-brane embedding functions $X^i(\sigma)$ with
$\sigma = (\sigma^0,\sigma^1,\ldots,\sigma^p)$, a Nambu Poisson tensor $\Pi^{iJ}
= \Pi^{i j_1 \ldots j_p}(X(\sigma))$, and volume fields $\eta_i = \eta_i(\sigma) d\sigma^1 \wedge \ldots \wedge d\sigma^p$ and $\tilde \eta_J
= \tilde\eta_{j_1 \ldots j_p} d\sigma^0$. (The ordered multi-index $J$ is defined as in \eqref{nambuian}.) Generalizing \eqref{etaPoissonsigma}, we introduce the \emph{Nambu sigma model} action
\begin{equation} \label{Nambu}
  S = \int_\Sigma \left( dX^i \wedge \eta_i + \tilde \eta_J \wedge d^pX^J  - \Pi^{iJ} \tilde\eta_J \wedge \eta_i
  - \frac{1}{2} G^{ij} \eta_i\wedge *\eta_j -\frac{1}{2} \tilde G^{IJ} \tilde \eta_I \wedge *\tilde \eta_J \right) \,,
\end{equation}
where $d^pX^J \equiv dX^{j_1} \wedge \ldots \wedge dX^{j_p}$. We notice the appearance of two types of metric fields: $G^{ij}$ 
and $\tilde G^{IJ}$
. For $p=1$ they are simply equal, but for $p>1$ their relation is much more intricate as we shall see. In principle, they could be independent.

The equations of motion of the \emph{topological Nambu sigma model}
\begin{equation} \label{topoNambu}
  S = \int_\Sigma \left( dX^i \wedge \eta_i + \tilde \eta_J \wedge d^pX^J  - \Pi^{iJ} \tilde\eta_J \wedge \eta_i\right)
\end{equation}
are
\begin{equation}
\begin{split}
&dX^i - \Pi^{iJ} \tilde\eta_J =0\,, \qquad d^pX^J - \Pi^{iJ} \eta_i = 0 \\
&d \eta_{i_k} - (-1)^k d\tilde\eta_I \, dX^{i_1} \wedge \ldots \wedge {d X}^{i_{k-1}} \wedge {d X}^{i_{k+1}}\wedge \ldots \wedge dX^{i_p}
+ \partial_{i_k} \Pi^{iJ} \,\tilde \eta_J \wedge \tilde \eta_i =0 \,.
\end{split}
\end{equation}
As in the $p=1$ case, consistency of the equations of motion is ensured by the fundamental identity \eqref{fundamental}, but the discussion is more involved and we shall not go into its details here. One of the motivations for the construction of the topological Nambu sigma model has been the problem of the quantization of  Nambu-Poisson structures. For this, the terms $\tilde \eta_J d^p X^J = \tilde \eta_{{j_1}\ldots {j_p}} dX^{j_1}\ldots dX^{j_p}$ in the action \eqref{topoNambu} need to be linearized (in the $dX$'s) with the help of further auxiliary fields to allow path integral quantization. Corresponding expressions can be found in \cite{Bouwknegt:2011vn}.

Let us now consider the full (non-topological) Nambu sigma model action \eqref{Nambu}. On shell, using the equations of motion for $\eta_i$ and $\tilde \eta_J$, it is equivalent to the following $p$-brane action
\begin{equation} \label{pbrane}
S' = -\int_\Sigma \left(\frac{1}{2} g_{ij} dX^i \wedge *dX^j
+ \frac{1}{2} \tilde g_{IJ} d^p X^I \wedge *d^p X^J
+ B_{iJ} d_0X^i \wedge d^p X^J\right)\,,
\end{equation}
where $d_0 X^i \equiv \left(\partial_0 X^i(\sigma)\right) d\sigma^0$.
The metric fields $g$, $\tilde g$ and the $1+p$ form background field $B$ are related to $G$, $\tilde G$ and $\Pi$ by matrix
relations
\begin{align} \label{open-closed-membrane}
G &= g  + B \tilde g^{-1} B^T \,, \quad \tilde G = \tilde g  + B^T g^{-1} B^T \,, \nonumber \\
\Pi &= - G^{-1} B \tilde g^{-1} = - g^{-1} B \tilde G^{-1}
\end{align}
that generalize the open-closed string relations \eqref{close-open-string} to the case of $p$-branes. (Note that $B$ and $\Pi$ are rectangular matrices.) It is possible to include an additional $1+p$ form field $\Phi$ that offers more freedom in the description, but for the sake of simplicity of presentation we shall refrain here from doing so.
Choosing for $\tilde g$ the anti-symmetrized product of $p$ copies of the metric $g$, the action \eqref{pbrane}
becomes
\begin{equation}
S'= \frac{1}{2} \int_\Sigma d^{p+1}\sigma \: \Big[ g_{ij} \partial_0 X^i \partial_0 X^j - \det(g_{ij} \partial_a X^i \partial_b X^j)\Big] - \int_\Sigma \frac{1}{(1+p)!} B_{i_0\ldots i_p} dX^{i_0} \wedge \ldots \wedge dX^{i_p}\,,
\end{equation}
which is related to the more familiar Nambu-Goto action featuring the pullback to the world volume of a $1+p$ form $B$-field background
\begin{equation}
S' = \int_\Sigma d^{p+1}\sigma \: \sqrt{\det \left(g_{ij} \partial_\alpha X^i \partial_\beta X^j\right)}
- \int_\Sigma X^* (B)\,.
\end{equation}

\section{Nambu-Dirac-Born-Infeld action}

An interesting application of the closed-open string relations is the Dirac-Born-Infeld (DBI) action, which is the effective action for open strings ending on a D-brane. Focusing just on the background fields $g$ and $B$, the action is
\begin{equation} \label{BIaction}
\frac{1}{g_s}\int d^nx \det{}^{1 \over 2} \left[ g + B\right]
= \frac{1}{g_s}\int d^nx \det{}^{1 \over 4} \left[ g \right] \det{}^{1 \over 4} \big[ \underbrace{g - Bg^{-1}B}_{G}\big] \,,
\end{equation}
where the second, more symmetrical expression, features the open string metric $G$.
Taking fluctuations into account, $B$ is replaced by $\mathcal F = B + F$ in \eqref{BIaction} and we obtain the full DBI action. Like  \eqref{BIaction}  this action has equivalent descriptions in terms of closed and open string parameters and fields, but the equivalence is more sophisticated: In the open string version, the commutative description with a background field $B$ is traded for a non-commutative description, where $F$ is promoted to a non-commutative field strength. Non-commutativity arises analogously to what happens in the Landau problem with a charged particle in a magnetic field.
A more careful analysis reveals that the approximate equivalence of commutative and non-commutative description is actually turned into an identity at the level the action, if we fall back to the semi-noncommutative level with Possion brackets replacing star commutators~\cite{Jurco:2001my}. The equivalent descriptions are\footnote{For clarity of presentation we have made some simplifying assumptions here, notably that $B$ and the couplings are constant and that $\Phi = 0$. Please see \cite{Jurco:2012yv} for expressions without these limiting assumptions.} \addtocounter{footnote}{-1}
\begin{equation}\label{DBI}
S_{DBI} = \int d^nx \frac{1}{g_s}\det{}^{1 \over 2} \left[ g + B + F\right]
= \int d^nx \frac{1}{G_s}\det{}^{1 \over 2} \left[ \hat G + \hat F\right] \,,
\end{equation}
where $\hat F$ is the semi-classical non-commutative field strength. The proof of the equivalence of the actions is based on the open-closed string relations \eqref{close-open-string} and a change of coordinates (semi-classical Seiberg-Witten map) $x^i \mapsto \rho^*(x^i) = \hat x^i =  x^i + \theta^{ij} \hat A_j$ induced by a transformation of Poisson structures  $\Pi \mapsto \Pi' = (1-\Pi F)^{-1}\Pi$. All hatted objects are obtained with the help of the map $\rho^*$ and transform covariantly under (semi-)non-commutative gauge transformations. The new coordinates $\hat x^i$ are called covariant coordinates and $\hat A_j$ is the ``semi-noncommutative'' gauge potential (see e.g. \cite{Jurco:2001kp}).

Trying to generalize \eqref{DBI} to $p>1$, i.e. to an effective description of (open) $p$-branes ending on another brane, we immediately face the problem that
$\det[g + B + F]$ makes no sense for square $g$ and rectangular $B$ and $F$. Fortunately, the second expression in \eqref{BIaction}  does lend itself to a $p>1$ generalization: $\det[g + B g^{-1}B^T]$ will become $\det[g + B\tilde g^{-1}B^T]$, where $\tilde g$ is the antisymmetrized product of $p$ copies of $g$.
All together the action must involve an integral density, but beyond $x + y = \frac{1}{2}$, the exponents $x$, $y$ in the following $F=0$ ansatz for the action
\begin{equation}
\frac{1}{g_m}\int d^nx \det{}^{x} \left[ \tilde g \right] \det{}^{y} \big[ \underbrace{g + B\tilde g^{-1}B^T}_{G}\big] \,,
\end{equation}
are so far undetermined. As in the string case we can add fluctuations
$B \mapsto \mathcal F = B + F$ and there is also a change of coordinates (Nambu-Poisson map) $x^i \mapsto \rho_N^*(x^i) = \hat x^i =  x^i + \Pi^{iJ} \hat A_J$ induced by a map between Nambu-Poisson tensors $\Pi \mapsto \Pi' = (1-\Pi\cdot F)^{-1}\Pi$.
The change of coordinates $\rho_N^*$ (Nambu SW map) is given by the flow of the vector field $\Pi(A) = \Pi^{iJ} a_J \partial_i$ with $F = dA$.
The corresponding full Nambu-Dirac-Born Infeld action
and its ``non-commutative'' counterpart are:\footnotemark
\begin{equation}
\begin{split}
S_{NDBI}
&= \int d^nx \frac{1}{g_m}\det{}^{x} \left[ \tilde g \right] \det{}^{y} \big[g + (B+F)\tilde g^{-1}(B+F)^T\big] \\
&= \int d^nx \frac{1}{G_m}\det{}^{x} \big[ \widehat{\tilde G} \big] \det{}^{y} \big[ \hat G + \hat F\widehat{\tilde G}^{-1} \hat F^T\big] \,.
\end{split}
\end{equation}
The proof of this {exact identity} is based on rather tricky matrix manipulations starting from the open-closed membrane relations \eqref{open-closed-membrane},
using the flow $\rho_N^*$, and appropriate relations for the ``membrane couplings''.
For the string case ($p=1$), a path-integral computation of the effective action gives $x = y = \frac{1}{4}$ \cite{Fradkin:1985ys}. For $p>1$ such a computation is an interesting open problem. Existence of equivalent commutative and non-commutative descriptions predict the exponents $x = \frac{p}{2(1+p)}$ and $y = \frac{1}{2(1+p)}$ \cite{Jurco:2012yv}. This prediction is confirmed by results of earlier computations based on $\kappa$-symmetry~\cite{Cederwall:1997gg}; for a review on super-$p$-branes with a comprehensive list of pertinent references, we refer to \cite{Sorokin:1999jx}.

\section{Summary}

We have presented a model that plays a similar role in the context of open $p$-branes as the Poisson sigma model and its non-topological generalizations do in the case of open strings. The model features a Nambu-Poisson tensor in place of the ordinary Poisson tensor. As a direct application we have derived a set of matrix relations that generalize the open-closed string relations to the case of $p$-branes. Based on these relations, we have derived a Dirac-Born-Infeld type action that enjoys an equivalence between a commutative description with background $p$-form field $B$ and a non-commutative (or rather `Nambuian') description without this field. Imposing this equivalence fixes the action essentially uniquely. We conjecture that the action thus obtained is the bosonic part of an all order effective action describing open $p$-branes ending on another brane (e.g.\ M2's ending on an M5 brane). Interestingly, new action does not feature the square root of the ordinary DBI action. The square root is rather found to be just a special case of a more general expression involving a different root.

\section*{Acknowledgements}
We would like to thank Jan Vysok\'y for helpful comments and the organizers of the Corfu Summer Institute 2011 for a memorable and productive conference.
The research of B.J. was supported by grant GA\v CR P201/12/G028.


\begin{thebibliography}{99}

\bibitem{Seiberg:1999vs}
  N.~Seiberg and E.~Witten,
  ``String theory and noncommutative geometry,''
  JHEP {\bf 9909}, 032 (1999)
  [hep-th/9908142].

\bibitem{Jurco:2001my}
  B.~Jurco, P.~Schupp and J.~Wess,
  ``NonAbelian noncommutative gauge theory via noncommutative extra dimensions,''
  Nucl.\ Phys.\ B {\bf 604}, 148 (2001)
  [hep-th/0102129].

\bibitem{Chen:2010br}
  C.~-H.~Chen, K.~Furuuchi, P.~-M.~Ho and T.~Takimi,
  ``More on the Nambu-Poisson M5-brane Theory: Scaling limit, background independence and an all order solution to the Seiberg-Witten map,''
  JHEP {\bf 1010}, 100 (2010)
  [arXiv:1006.5291 [hep-th]].

\bibitem{Jurco:2012yv}
  B.~Jurco and P.~Schupp,
  ``Nambu-Sigma model and effective membrane actions,''
  arXiv:1203.2910 [hep-th].

\bibitem{Nambu:1973qe}
  Y.~Nambu,
  ``Generalized Hamiltonian dynamics,''
  Phys.\ Rev.\ D {\bf 7}, 2405 (1973).

\bibitem{Takhtajan:1993vr}
  L.~Takhtajan,
  ``On Foundation of the generalized Nambu mechanics,''
  Commun.\ Math.\ Phys.\  {\bf 160}, 295 (1994)
  [hep-th/9301111].

\bibitem{Ikeda:1993fh}
  N.~Ikeda,
  ``Two-dimensional gravity and nonlinear gauge theory,''
  Annals Phys.\  {\bf 235}, 435 (1994)
  [hep-th/9312059].

\bibitem{Schaller:1994es}
  P.~Schaller and T.~Strobl,
  ``Poisson structure induced (topological) field theories,''
  Mod.\ Phys.\ Lett.\ A {\bf 9}, 3129 (1994)
  [hep-th/9405110].

\bibitem{Cattaneo:1999fm}
  A.~S.~Cattaneo and G.~Felder,
  ``A Path integral approach to the Kontsevich quantization formula,''
  Commun.\ Math.\ Phys.\  {\bf 212}, 591 (2000)
  [math/9902090].

\bibitem{Jurco:2001kp}
  B.~Jurco, P.~Schupp and J.~Wess,
  ``Noncommutative line bundle and Morita equivalence,''
  Lett.\ Math.\ Phys.\  {\bf 61}, 171 (2002)
  [hep-th/0106110].


\bibitem{Bouwknegt:2011vn}
  P.~Bouwknegt and B.~Jurco,
  ``AKSZ construction of topological open p-brane action and Nambu brackets,''
  arXiv:1110.0134 [math-ph].

\bibitem{Fradkin:1985ys}
  E.~S.~Fradkin and A.~A.~Tseytlin,
  ``Quantum String Theory Effective Action,''
  Nucl.\ Phys.\ B {\bf 261}, 1 (1985).

\bibitem{Cederwall:1997gg}
  M.~Cederwall, B.~E.~W.~Nilsson and P.~Sundell,
  ``An Action for the superfive-brane in D = 11 supergravity,''
  JHEP {\bf 9804}, 007 (1998)
  [hep-th/9712059].

\bibitem{Sorokin:1999jx}
  D.~P.~Sorokin,
  ``Superbranes and superembeddings,''
  Phys.\ Rept.\  {\bf 329}, 1 (2000)
  [hep-th/9906142].

\end{thebibliography}
\end{document}